# A CFD Model for Heat and Mass Transfer Leading To Plume Formation within Wet Cooling Towers

**Favre Luc[1], Ferrand Martin[1,2]**
[1]Centre d'Enseignement et de Recherche en Environnement Atmosphérique
École des Ponts ParisTech, 6/8 Avenue Blaise Pascal, Champs-sur-Marne, France
luc.favre@enpc.fr / favre.luc05@gmail.com ; martin.ferrand@edf.fr
[2]MFEE, EDF R&D
6 Quai Watier, Chatou, France

***Abstract*** - The crucial role played by Wet Cooling Towers (WCT) in many electricity production plants (e.g. nuclear power plants) make them a key parameter in the industrial design of such facilities. Their impact over the cooling water consumption and surrounding atmosphere through the formation and dispersion of a humid air plume has pushed the need to obtain proper models and simulations in order to anticipate those effects. In this work, we tackle this issue through a dedicated modelling in the CFD solver code_saturne. Specific modeling includes heat and mass transfer (convection and evaporation) between the injected water and the air flow that are validated against experimental results obtained in a reduced scale WCT experimental loop. Satisfying agreement is obtained for several parameters such as air and water exit temperatures, evaporation mass flow rate and total exchanged thermal power. This constitutes an important first step for detailed CFD predictions of WCT water consumption and humid air plume atmospheric dispersion.

## 1. Introduction

Cooling towers are nowadays a key component of many power plants and usually considered as useful means to achieve the refrigeration required for their operation. In the case of Natural Draft Wet Cooling Towers (NDWCT), the hot water to be cooled (usually coming out of a condenser) is sprayed inside the tower in which a buoyancy-driven cold air flow from the outside is created using the density difference between the top and the bottom of the cooling tower. The injected water is cooled both through convection and evaporation in interaction with the cold air, making it a two-phase heat and mass transfer medium.

In order to improve the efficiency of those heat and mass transfers, cooling towers are equipped with so-called fill-packing zones which are usually composed of a dense accumulation of plastic layers whose role is to spread the injected water into liquid films flowing down the packing to increase the exchange surface with the cold air. Below the packing zone, water falls down the tower under the form of a rain before reaching the ground where it is collected before being re-injected in the cooling circuit.

The physics at stake within cooling towers is thus intrinsically complicated since it includes multi-phase flows, turbulence, phase change (evaporation) and heat transfer. Therefore, achieving detailed modelling of cooling towers through scaling or one-dimensional approaches is very difficult, if not infeasible. That is why 3D approaches using CFD are increasingly considered for many reasons [1-5]:

• Capturing 3D heterogeneity in the flow quantities (temperature, humidity, velocity, etc.) ;
• Reproducing detailed geometries of cooling towers and their impact on the overall flow and heat transfer ;
• Studying the impact of outer atmospheric conditions (stability, temperature, crosswind) on the thermal performance of the cooling tower ;
• Implementing micro-physics modelling both for multi-phase transfers and atmospheric flows.

Recent insights of detailed cooling towers local physics are starting to propose finer modelling to be applied in future simulations, as in the work of Jourdan et al. who studied the local hydrodynamics of water films in the packing zones [6, 7].

In this work, we present first steps of development in code_saturne CFD solver [8] to model WCT physics and associated validation based on a large-scale experiment handled by EDF at Bugey nuclear power plant.

## 2. CFD modelling
### 2.1. Equations
In order to study the air flow, we first solve the Navier-Stokes equations (Eq. 1 and 2) for the humid air mixture (that includes dry air, water vapour and liquid water condensate when the air becomes saturated in humidity).

$$\frac{\partial \rho_h}{\partial t} + \text{div}\,(\rho_h \boldsymbol{u}) = \Gamma \qquad (1)$$

$$\rho_h \frac{\partial \boldsymbol{u}}{\partial t} + \underline{\underline{\textbf{grad}}}\,(\boldsymbol{u}) \cdot (\rho_h \boldsymbol{u}) = -\,\textbf{grad}\,(P^*) + \textbf{div}\,\left(2\,(\mu_h + \mu_T)\,\underline{\underline{\boldsymbol{S}}}^D\right) - \textbf{div}\,\left(\rho_h \underline{\underline{\boldsymbol{R}}}^D + 2\mu_T \underline{\underline{\boldsymbol{S}}}^D\right) \\ + (\rho_h - \rho_0)\,\boldsymbol{g} + \boldsymbol{ST_u} - \rho_h \underline{\underline{\boldsymbol{K}}} \cdot \boldsymbol{u} + \Gamma\,(\boldsymbol{u}^{in} - \boldsymbol{u}) \qquad (2)$$

Where $\rho$ is the density, **u** the velocity, P* the pressure, **S** and **R** the strain rate and Reynolds Stress tensor, **g** the gravity, $\mu$ the viscosity, **K** a head loss tensor and $\Gamma$ the mass source term representing the increase of the bulk mass due to injected water evaporation. Subscript h and T denote humid air and turbulent properties, superscript D means deviatoric part of a tensor. Turbulence is modelled using the k-$\varepsilon$ model with linear production [9].

The energy conservation is also solved for the humid air temperature T (Eq. 3).

$$c_{p,h}\left[\frac{\partial \rho_h T}{\partial t} + \text{div}\,(\rho_h T \boldsymbol{u})\right] = \text{div}\,(\lambda_h \textbf{grad}\,(T)) + ST_T \qquad (3)$$

Where cp is the specific heat capacity and $\lambda$ the thermal conductivity.

For water quantities, we solve scalar transport equations for the water mass fraction in air $y_w$ and the mass of injected liquid water per kg of humid air $Y_{l,r}$. The injected water components are transported with a drift velocity [10] versus the humid air to follow the liquid flow in the packing and the rainfall down the tower.

### 2.2. Closure laws
The evaporation is modelled using the formulation of Poppe [11] with an evaporation mass flux in the fill pack zone computed following Eq. 4.

$$\beta_x a_i = AF_e \left(\frac{F_a}{F_e}\right)^n \qquad (4)$$

Where $\beta_x$ is the evaporation coefficient, $a_i$ the interfacial area concentration, $F_a$ and $F_e$ the dry air and injected water surface mass flow rates respectively. A$\approx$0.8 and n$\approx$0.6 are adjusted coefficients that represent the evaporative capacity of a given fill pack structure.

The model is closed by relating the convection heat transfer to the evaporation mass transfer through Bosnjakovic formulation of the Lewis factor [12]. In the rain zone, the interfacial Nusselt number is computed following Ranz & Marshall correlation [13]. Finally, the friction between air and rain is modelled following the approach of Dreyer [14].

## 3. Validation
### 3.1. The MISTRAL experiment

The MISTRAL loop is an experimental setup dedicated to cooling towers fill pack study and is located at Bugey nuclear power plant in France. Its goal is to perform measurements and analyses of fill packs thermal performances before their implementation in operating or future NDWCT. It consists of a mechanically induced draft facility using a fan located at the outlet tower at an 18.7 m height. Below the outlet tower is a square section of 7 × 7 m2 and 6.6 m height (starting 10 m above the ground level) designed to contain :
- The fill pack to be tested ;
- The hot water sprays, designed to control both injected water temperature and flow rate, located 1 m above the fill pack ;
- The droplets scrubbers, whose role is to block large droplets formed in the humid air during the evaporation process.

Below the packing zone, the cooled water falls under the form of a cold rain before being collected in a basin, in which cold water temperature is measured to quantify the cooling capacity of the studied fill pack. The total rainfall is 10 m high, which is close to real NDWCT. The air inlet is located at the east part of the facility, with an inlet section of 5 m width, 10 m height and 25.5 m distance to the centre of the packing zone.

The MISTRAL bench is equipped with several sensors that provide with relevant measurements such as:
- Pressure drop through the packing zone, the eliminators and the entire section ;
- Air temperature at inlet and outlet ;
- Water temperature at injection and in the collection basin ;
- Atmospheric pressure and humidity of outside air draught into the facility ;
- Air and water mass flow rates.

Those series of measurements allow to precisely calculate important quantities such as the evaporated mass flow rate, dissipated thermal power on the water side, and the received thermal power on the air side. Comparison between the energy balances of both fluids usually results in a difference less than 5%.

### 3.2. Simulation results

The MISTRAL experiment is simulated using code_saturne and the presented modelling to validate its formulation. Fig. 1 presents a typical result of a 2D MISTRAL bench simulation. We can see that the air flow coming from the east and exiting at the top presents two recirculation zones: one in the room situated at the west of the building and another one at the bottom of the rain zone, resulting from the interfacial friction between humid air and rain drops. Most of the air heating is achieved in the fill pack zone.

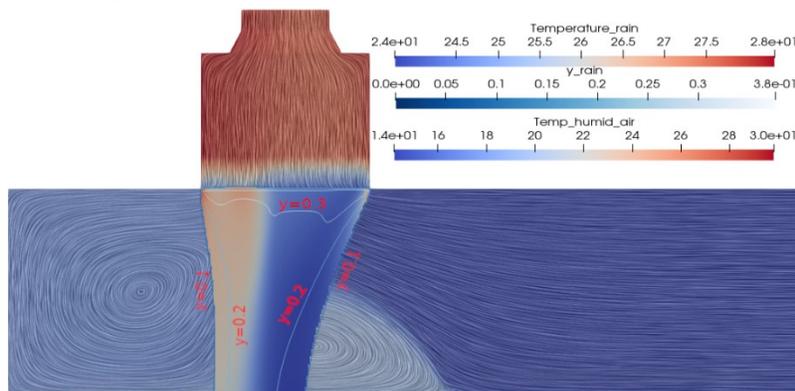

Fig. 1: MISTRAL case simulation results, right corresponds to east direction. Inlet is located at the right. Fill pack zone is located above the rain zone.

Fig. 2 compares obtained code_saturne results with the experimental measurements. A satisfying agreement is obtained for the rain and humid air exit temperatures for the whole range of water and air mass flow rates. A stable overestimation of the water temperature and underestimation of the air temperature are observed which is in accordance with the constant underestimation of the evaporated water mass flow. This could be explained by the fact that we do not simulate the rain injection zone before the packing yet, where hot water would evaporate quicker than in the cold rain zone. Moreover, we did not conduct an optimization of the packing evaporation correlation as in other studies [3]. Globally, we achieve temperature predictions within a 10% error margin and 15% for the evaporation rate.

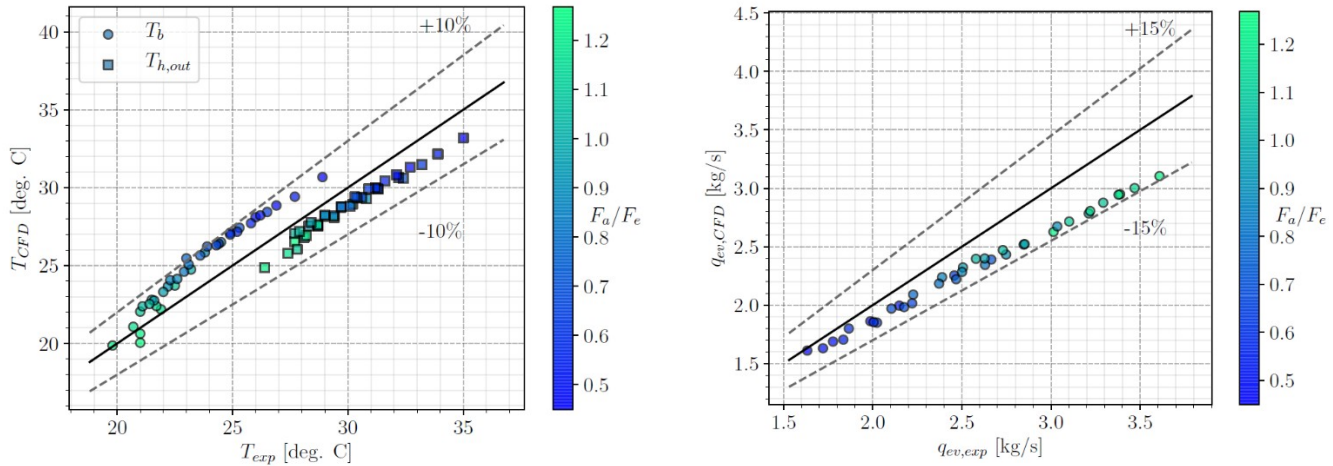

Fig. 2: CFD and experiments comparisons of air and rain exit temperatures (left) and evaporated mass flow rates (right) for 55 MISTRAL cases.

## 4. Conclusion

All in all, the proposed modelling for WCT enables reproducing satisfactory results of importance to predict the water consumption along with the humid air plume characteristics that will be released in the atmosphere. Future work will focus on simulating the natural draft in WCT together with the ambient atmospheric flow to estimate the plume dispersion in nearby areas.

## Acknowledgements

This work is supported by EDF and Ecole des Ponts ParisTech.